\documentclass[pre,twocolumn,notitlepage,superscriptaddress,showpacs]{revtex4-2}
\usepackage{amsmath,amsfonts,amssymb,comment}
\usepackage[pdftex]{graphicx}
\usepackage[dvipsnames]{xcolor}
\definecolor{myDarkRed}{rgb}{0.6, 0, 0}

\begin{document}
\title{Localized active phase in the coevolving voter model with fixed links}
\author{Hoseung Jang}
\email{jhs3349@gmail.com} 
\affiliation{Advanced-Basic-Convergence Research Institute, Chungbuk National University, Cheongju, Chungbuk 28644, Korea}
\author{Maxi San Miguel}
\email{maxi@ifisc.uib-csic.es}
\affiliation{Institute for Cross-Disciplinary Physics and Complex Systems (IFISC), UIB-CSIC, 07122 Palma de Mallorca, Spain}
\author{Byungjoon Min}
\email{bmin@cbnu.ac.kr}
\affiliation{Advanced-Basic-Convergence Research Institute, Chungbuk National University, Cheongju, Chungbuk 28644, Korea}
\affiliation{Department of Physics, Chungbuk National University, Cheongju, Chungbuk 28644, Korea}

\date{\today}\date{\today}

\begin{abstract}

The coevolving voter model (CVM) describes the evolution of node states 
and network topology through copying and rewiring dynamics. 
It traditionally assumes that all ties can be rewired, overlooking 
unbreakable real-world relationships like family or close friendships. 
To address this, we introduce fixed links that cannot be rewired and study 
how these structural constraints modify the collective dynamics of the CVM.
In addition to the active and frozen phases of the CVM, we find a new localized active phase in which activity is sustained, beyond a percolation transition, by a giant component of fixed links. The existence of this new phase is predicted by a moment-closure approximation that goes beyond the standard pair approximation. The finite-size manifestation of this phase is an asymmetric fragmented phase. In addition, the active-frozen transition of the CVM, manifested as a fragmentation transition for finite systems,  is delayed by the existence of fixed links to larger values of the rewiring rate. 
\end{abstract}
\maketitle

\section{Introduction}

From the polarization of opinions to the spread of infectious diseases, 
the collective phenomena observed in complex systems are 
governed by underlying networks \cite{jensen2022complexity,newman2018networks}. 
By representing these systems as networks where nodes denote individual
components and edges represent their interactions, we can discover 
how structural topologies shape the time evolution of the system's states
\cite{newman2018networks,boccaletti2006complex}. 
However, in most real-world scenarios, the connections between 
individuals  are rarely static; instead, the underlying network structure 
evolves over time \cite{zimmermann2004coevolution,holme2012temporal,barabasi1999emergence,masuda2016guide,jang2026feedback}.
Moreover, in many complex systems, the structure of the network
itself continuously changes over time in response to the dynamic states 
of their elements \cite{holovatch2017complex,gross2008adaptive,berner2023adaptive}.
This coupled dynamics of node states and network structure provides an essential layer of complexity in understanding
various phenomena, including opinion formation 
\cite{holme2006nonequilibrium,vazquez2008generic,saeedian2020absorbing,liu2023emergence}, 
epidemic dynamics \cite{gross2006epidemic,marceau2010adaptive}, spin dynamics \cite{biely2009socio}, 
complex contagion \cite{min2023threshold,burgio2025characteristic}, and game-theoretical models \cite{gonzalez2023network}.
Such intertwined dynamics give rise to complex phenomena in both 
the network structure and node states that do not emerge in static networks \cite{berner2023adaptive}.

The coevolving voter model (CVM) provides a paradigmatic framework 
for investigating the 
consequences of the co-evolution of node states and network topology~\cite{holme2006nonequilibrium,vazquez2008generic,durrett2012graph}.
The standard voter model (VM) \cite{clifford1973model,liggett1999stochastic} has
been extensively studied with a focus on non-equilibrium ordering dynamics 
and consensus formation \cite{redner2019reality,castellano2003incomplete,sood2005voter,suchecki2004conservation,suchecki2005voter,vazquez2008analytical,kim2025competition}. 
While the dynamics of the classical VM relies solely on nodes copying
the opinions of their neighbors on a static network, 
real-world interactions may require a dynamic adaptation. 
In the CVM \cite{holme2006nonequilibrium,vazquez2008generic}, the network structure continuously adapts to the changing 
states of its constituents.
The original CVM consists of two dynamic processes: copying and rewiring. 
Each node, in one of two states, can either copy 
the state of a randomly selected neighbor or rewire its connection if the
neighbor is in a different state \cite{vazquez2008generic}. 
The relative rate of link rewiring to copying is determined by 
the network plasticity $p$. 

A key feature of the CVM is an absorbing phase transition between 
a dynamically active phase and a frozen phase, characterized by the
density $\rho$ of active links, that is, links connecting nodes in different states
\cite{vazquez2008generic}. As the rewiring probability $p$ increases,
the active-link density decreases and vanishes at the critical point $p_c$. 
In a finite system, the active phase eventually reaches a consensus 
state due to finite-size fluctuations. 
In contrast, the frozen phase consists of fragmented components, 
each of which maintains its own internal consensus.
Several extensions of the CVM have been studied, including multilayer
coevolution \cite{diakonova2014absorbing}, non-linear interactions \cite{min2017fragmentation,raducha2018coevolving,min2019multilayer},
aging \cite{min2025aging}, and higher-order interactions \cite{horstmeyer2020adaptive,papanikolaou2022consensus,golovin2024polyadic}, to name a few.

Although the CVM successfully captures the adaptive nature of social ties 
through rewiring, it often overlooks unbreakable connections  \cite{kobayashi2019structured}.
Certain social ties, such as family or long-term friendships, tend to 
persist or remain permanent regardless of conflicts between them.
To bridge this gap, we incorporate ``fixed links'' in the CVM, where a specific
fraction of links is preserved and immune to rewiring, independent of the nodes' states. 
These fixed links are analogous to the zealots considered in the ordinary voter model \cite{galam1997,mobilia2003,mobilia2015nonlinear,khalil2018}. Zealots are nodes that never change state. The density of zealots is a parameter driving transitions between opinion coexistence and consensus in different voter models \cite{kitching2025breaking}. 
In addition, the fixed links can be regarded as a limiting case of link aging in the CVM, in which certain links become permanently resistant to rewiring \cite{min2025aging}.
A natural question in this context is how the density $q$ of fixed links modifies the phases found in a CVM. We find that in addition to the active and frozen phases of the CVM, a new localized active phase appears in the ($p,q$) parameter space of the CVM with fixed links. While active and frozen phases manifest themselves in finite systems as consensus and fragmented phases, the localized active phase in a finite system becomes an asymmetric fragmented phase. Fixed links create a quenched structural constraint in the network, so that when $q$ exceeds the percolation threshold, a giant fixed-link network component emerges that supports a localized active state of opinion coexistence.

The remainder of this paper is organized as follows. 
In Sec.~II, we introduce the coevolving voter model with fixed links. 
In Sec.~III, we develop the theoretical approaches, including a 
moment-closure approximation for the absorbing transition and a 
percolation analysis of the subgraph formed by fixed links. 
In Sec.~IV, we present the numerical results and characterize the 
resulting dynamical regimes. Finally, in Sec.~V, we summarize and 
discuss our main findings.

\section{Coevolving voter model with fixed links}

\begin{figure}[t]
\centering
\includegraphics[width=\linewidth]{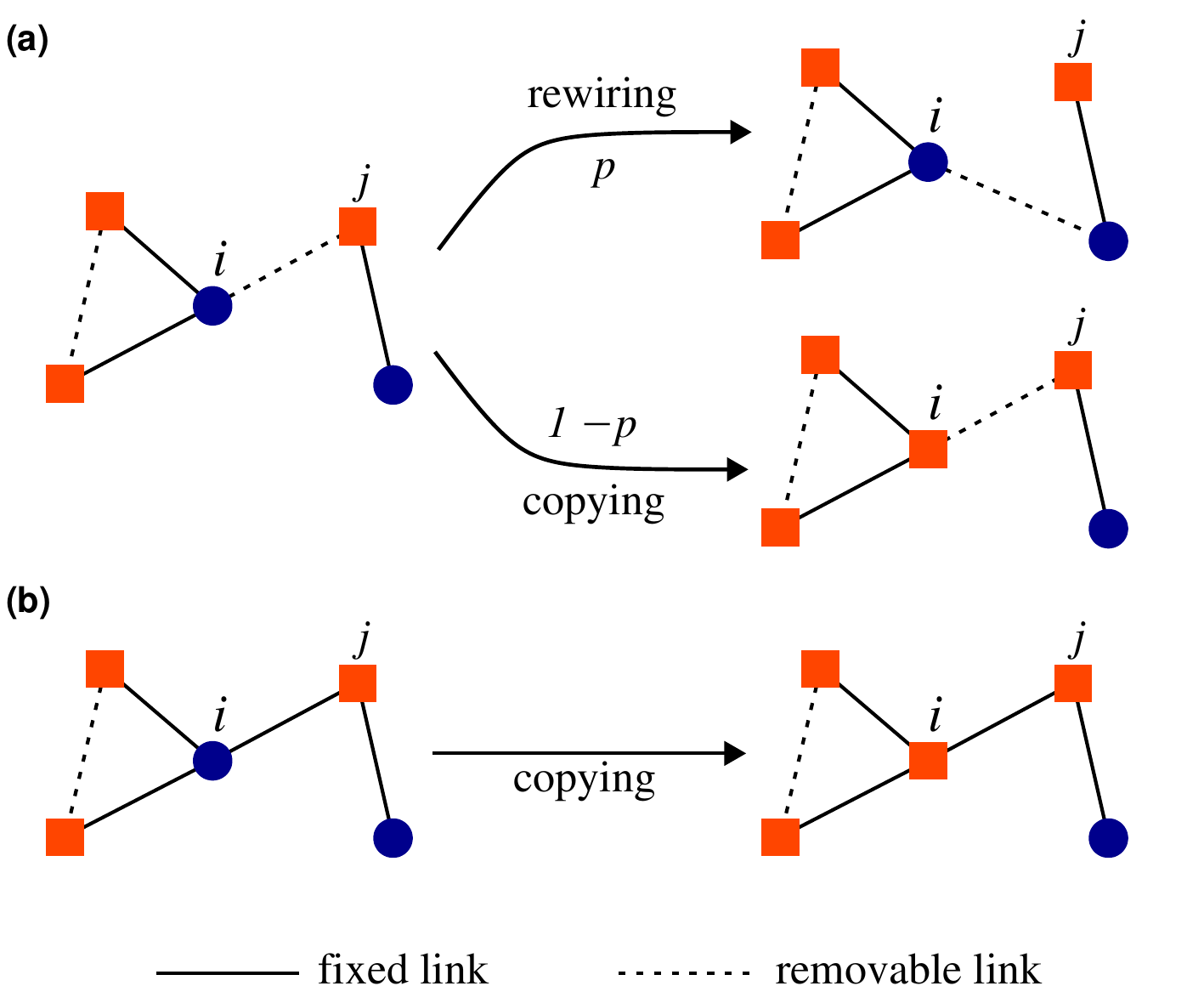}
\caption{\label{fig_model} 
Coevolving voter model with fixed links.
Schematic illustration of the updating rule when node $i$ and node $j$ are connected via (a) a removable link and (b) a fixed link.  
Solid lines represent fixed links and dotted lines represent removable links.
}
\end{figure}

We consider a network of $N$ nodes. Each node $i$ can be in one of two states, i.e., 
$\sigma_i \in \{+1, -1\}$. In this work, we use random regular networks with degree $z$ as an initial condition.
The network consists of two types of edges: 
fixed links, which represent permanent social bonds (e.g., family), and 
removable links, which represent casual acquaintances. 
We denote the fraction of fixed links as $q$, while the remaining $1-q$ fraction
of links are subject to a rewiring process. 
Since fixed links cannot be removed from the network, they introduce
a form of quenched structural disorder.
The fixed-link density $q$ therefore controls the degree of structural constraint,
with $q=0$ corresponding to the classical CVM \cite{holme2006nonequilibrium,vazquez2008generic} and $q=1$ corresponding
to the ordinary voter model on a completely static network.

The dynamical rules of the CVM with fixed links are as follows (see Fig.~\ref{fig_model}).
At each time step, a node $i$ is randomly selected and one of its neighbors $j$ 
is also selected uniformly at random. If their states are different, i.e.,
$\sigma_i \neq \sigma_j$, the system evolves according to the following rules: 
i) if the chosen link connecting nodes $i$ and $j$ is fixed (unremovable),
node $i$ adopts the state of node $j$ with probability one, and
ii) if the link between $i$ and $j$ is a removable link,
the connection is removed with probability $p$, and node $i$ establishes a new link
with a randomly chosen node $l$ that shares the same state
($\sigma_i = \sigma_l$). 
With probability $1-p$, node $i$ copies the state of its neighbor $j$, 
such that $\sigma_i$ becomes equal to $\sigma_j$.
The competition between the rewiring with probability $p$ and the copying with probability 
$1-p$ governs the network evolution and opinion formation of the system. 
We repeat the procedure until reaching a steady state.

\section{Theoretical approaches}

In this section, we explore the impact of fixed links from two different perspectives. First, we examine the structural effect of fixed links focusing on the formation of a giant connected component composed of fixed links. Second, we study the dynamical effect of fixed links using moment-closure approximations.

\subsection{Percolation of fixed links}

We consider the connected component composed solely of fixed links 
since it provides a permanent connected structure. 
Each fixed link is chosen with probability $q$ and hence
this subgraph corresponds to ordinary bond percolation with bond
occupation probability $q$ on the underlying network.
We then use the conventional technique for identifying 
the size of the giant component based on the generating functions.
For a random network with degree distribution $P(k)$, we define the
degree and excess-degree generating functions as \cite{newman2018networks}
\begin{align}
G_0(x) = \sum_{k=0}^{\infty} P(k)x^k,\quad 
G_1(x) = \frac{G_0'(x)}{G_0'(1)},
\end{align}
where $G_0'(x)=dG_0(x)/dx$.
Let $u$ denote the probability that by following a randomly chosen
link, one does not reach the giant component through fixed links.
The probability $u$ is given by the following self-consistency equation:
\begin{align}
u = 1-q+qG_1(u).
\label{eq:u_fixed}
\end{align}
Once we obtain the probability $u$ by solving iteratively Eq.~(\ref{eq:u_fixed}),
the size of the giant component formed by fixed links can be obtained as 
\begin{align}
S_{\rm f}=1-G_0(u),
\label{eq:Sfixed}
\end{align}
where the size is normalized to 1.

For a random regular network of degree $z$,
$G_0(x)=x^z$ and $G_1(x)=x^{z-1}$, so that
\begin{align}
u &= 1-q+qu^{z-1},\\
S_{\rm f} &= 1-u^z.
\end{align}
The emergence of the giant component composed of fixed links is depicted
in the inset of Fig.~\ref{fig_phase}.
The giant component emerges when the trivial solution $u=1$ loses its stability, which gives $q_c G_1'(1)=1$.
Therefore, for a random regular network, the percolation threshold is \cite{callaway2000network,cohen2000resilience}
\begin{align}
q_c=\frac{1}{z-1}.
\label{eq:qc}
\end{align}
For $q>q_c$, the fixed-link subgraph contains a giant connected
component that cannot be fragmented by rewiring.
This giant component allows active links to persist even under strong rewiring.

\subsection{Moment-closure approximations}

To obtain analytical insight of the density $\rho$ of active links, we develop a moment-closure approximation 
\cite{kimura2008coevolutionary,demirel2014moment}. 
We denote by $[AB]$ the number of active links per node, i.e., links connecting nodes in different states,
where $A$ and $B$ represent the two different states. 
For a random regular network of degree $z$, the density $\rho$ of active
links is related to $[AB]$ as
\begin{align}
\rho = \frac{2[AB]}{z}.
\end{align}
Since rewiring occurs only when the selected active link is
removable, its effective probability is $p(1-q)$, whereas a copying
event occurs with probability $q+(1-q)(1-p)$. 
Thus, the evolution of the active-link density is then given by
\begin{align}
\frac{d[AB]}{dt}=&\frac{2}{z}\Big\{ (1-p+pq) \big[(z-2)[AB]-4[ABA]\big] \\
&- p(1-q)[AB] \Big\} \nonumber
\label{eq:AB}
\end{align}
where $[ABA]$ is the density of triplets in which the central
node is connected to two nodes in the opposite state. Here, we 
use the symmetry between the two states, $[ABA]=[BAB]$, and 
hence combine the contributions of the two types of triplets into $[ABA]$.

The pair approximation \cite{vazquez2008generic,durrett2012graph} is recovered when the triplet density  $[ABA]$ is approximated as: 
\begin{align}
[ABA] \approx \frac{z-1}{z}\frac{[AB]^2}{2[B]},
\end{align} 
where $[B]$ is the fraction of nodes in state $B$.
Here, we use $[B]=1/2$ for the symmetric initial condition~\cite{demirel2014moment}. 
The equation for the active-link density becomes 
\begin{align}
\frac{d[AB]}{dt}=&\frac{2[AB]}{z}\Big\{ (1-p+pq) \big[(z-2)-4 \frac{z-1}{z}[AB] \big] \nonumber \\
&- p(1-q) \Big\}. 
\label{eq:pair}
\end{align}
Note that Eq.~(\ref{eq:pair}) for the CVM with fixed links is identical to the 
equation for the pair approximation for the ordinary CVM with $\tilde{p}=p(1-q)$ 
as an effective plasticity \cite{vazquez2008generic}.
The transition point $p_{c,PA}$ for the absorbing phase transition is given by
\begin{align}
p_{c,PA}= \left( \frac{1}{1-q} \right) \frac{z-2}{z-1},
\end{align}
for random regular networks.

In order to obtain a more accurate approximation,
we next consider the time evolution of $[ABA]$ to incorporate 
triplet correlations. 
The corresponding moment equation involves several quadruplet motifs.
We close this hierarchy using a second-order moment-closure approximation \cite{demirel2014moment,kimura2008coevolutionary},
in which quadruplet densities are expressed in terms of pair and triplet densities.
The time evolution of $[ABA]$ derived from the moment-closure approximation 
is given by
\begin{widetext}
\begin{align}
\frac{d[ABA]}{dt}=&\frac{1}{z}\Bigg\{[1-p(1-q)]\Bigg[-4[ABA]+\frac{2\big[(z-1)[AB]-2[ABA]\big]^2}
{z-2[AB]}
\nonumber\\
&+\frac{z-2}{2(z-1)}
\frac{\big[(z-1)[AB]-2[ABA]\big]^2}{[AB]}-\frac{2(3z-4)}{z-1}\frac{[ABA]^2}{[AB]}\Bigg]
-4p(1-q)[ABA]
\Bigg\}.
\label{app:dABA_closed}
\end{align}
\end{widetext}
The full derivation of the moment-closure approximations
is given in Appendix~\ref{app:moment_closure}.

The steady-state solution $\rho^*$ of the density of active links 
for $d[AB]/dt = d[ABA]/dt=0$ is given as 
\begin{align}
\rho^*=\frac{(z-1)^2 \tilde{p}^{2}-2(z^2-2z+2)\tilde{p}+z(z-2)}
{2\left[(z-1)^2 \tilde{p}^{2}-(2z^2-3z+2)\tilde{p}+z(z-1)\right]},
\label{eq:rho_theory}
\end{align}
where $\tilde{p}=p(1-q)$ is the effective plasticity.
The absorbing transition from the active to frozen phase occurs 
when the nontrivial stationary solution vanishes, $\rho^*=0$. 
Therefore, the transition point is predicted as:
\begin{align}
p_c(q)=\left(\frac{1}{1-q}\right)\frac{\sqrt{3z^2-6z+4}-2}{\sqrt{3z^2-6z+4}+1}.
\label{eq:pc}
\end{align}

Equation~(\ref{eq:pc}) provides an approximate prediction of the
transition line \cite{kimura2008coevolutionary,demirel2014moment}, 
since the moment-closure scheme neglects higher-order
correlations and treats the quenched fixed-link disorder 
in an annealed and homogeneous manner. 
However, it qualitatively captures the absorbing phase transition,
with $p_c$ increasing as $q$ increases.
While the moment-closure approximation captures the dynamical effect of fixed links at the annealed level, the percolation analysis accounts for their quenched structural effect.
These two approaches together provide a theoretical framework for understanding the macroscopic states.

\subsection{Phase diagram}

\begin{figure}
\centering
\includegraphics[width=\linewidth]{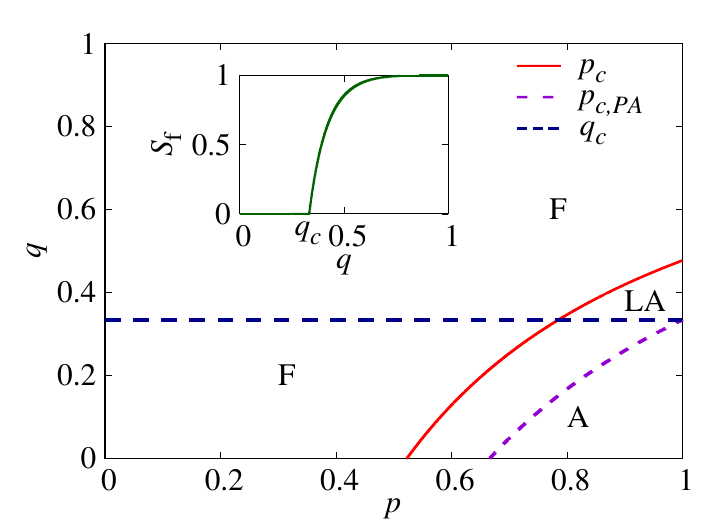}
\caption{\label{fig_phase}
Phase diagram from the absorbing transition lines predicted by the triplet approximation $p_c$ and the pair approximation $p_{c, PA}$, together with the percolation threshold $q_c$ of the fixed-link subgraph, for random regular networks with $z=4$.
Here, $\mathrm{A}$, $\mathrm{F}$, and $\mathrm{LA}$ denote active, frozen, and localized active phases.
The inset shows the size $S_{\rm f}$ of the giant component composed of fixed links 
as a function of $q$, as predicted by Eq.~(5).
}
\end{figure}

Figure~\ref{fig_phase} shows the phase diagram of the CVM with fixed 
links in the $(p,q)$ plane, predicted by the pair and moment-closure approximations 
and the percolation threshold of fixed links. 
The phase diagram is characterized by two distinct thresholds: 
the first is the absorbing transition line $p_c(q)$, which originates 
from the competition between copying and rewiring dynamics.
The second is the percolation threshold $q_c=1/(z-1)$ of the subgraph 
composed solely of fixed links. 
While $p_c(q)$ represents the dynamical suppression of activity by rewiring, 
the percolation threshold $q_c$ corresponds to the emergence of an extensive permanent backbone that cannot be fragmented by the rewiring process.

For $p<p_c(q)$, copying is sufficiently frequent relative to the rewiring.
In this regime, the system remains in the active phase regardless of
whether $q$ is below or above the percolation threshold $q_c$.
For $p>p_c(q)$, the behavior depends on whether the fixed links
percolate. When $q<q_c$, the fixed links form only finite components,
and strong rewiring drives the system into the frozen phase.
In contrast, when $q>q_c$, the fixed links form a giant connected
component that remains intact under rewiring. Although activity outside
this backbone is strongly suppressed and the surrounding network becomes
nearly frozen, active links can persist within the giant component
because nodes in different states remain connected through fixed links.
This gives rise to a localized active phase where the active links are  
concentrated mainly within the fixed-link giant component. 

The localized active phase is a new phase created by the existence of fixed links and that does not exist in the CVM. We also  note that while the transition line 
$p_{c,PA}$ based on the pair approximation predicts
the existence of the active and frozen phases of
the CVM with fixed links, it fails in predicting the localized
active phase since the transition line meets the percolation threshold at $p=1$.

\section{Numerical Results}

\begin{figure*}[t]
\centering
\includegraphics[width=\linewidth]{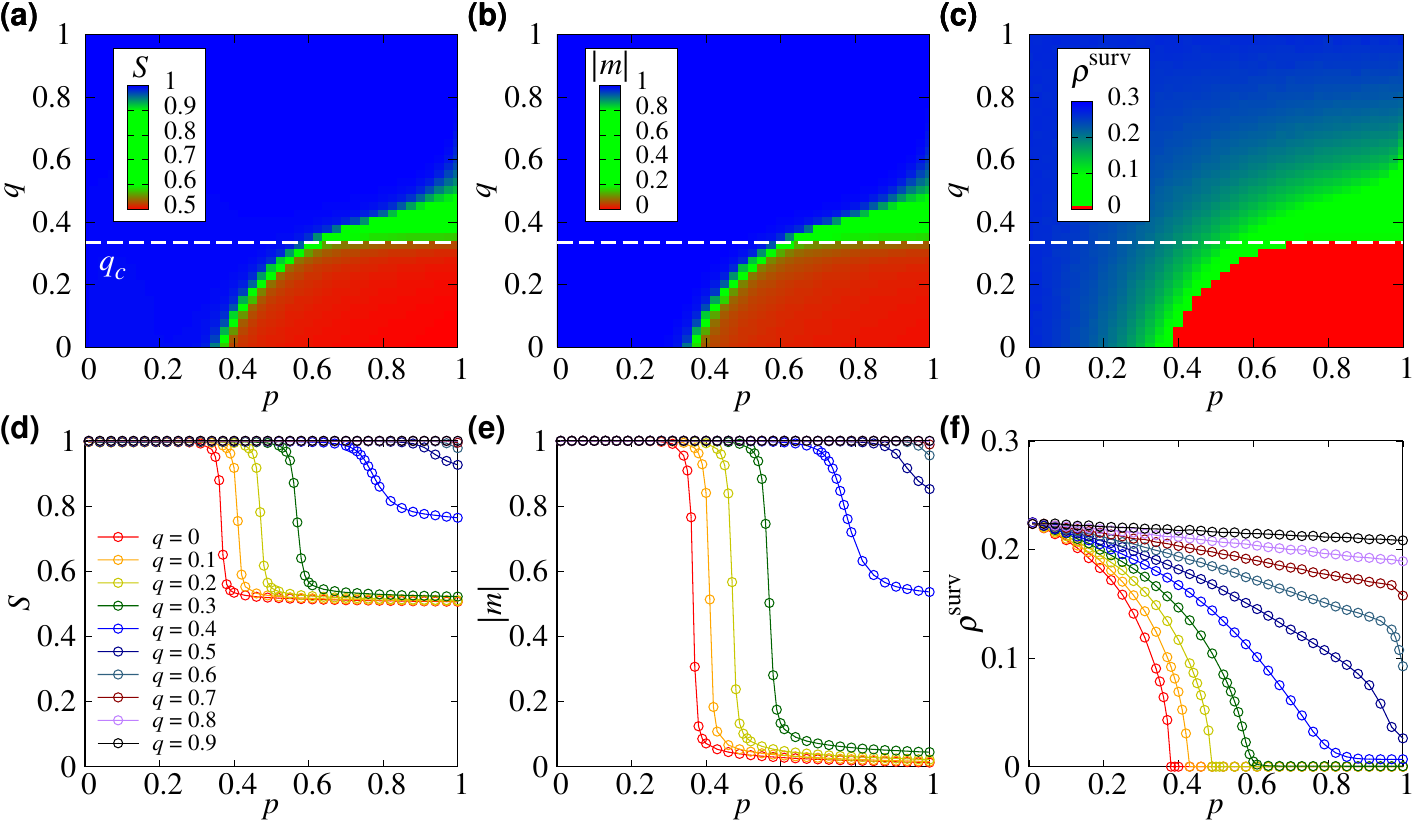}
\caption{\label{fig_RR}
Color maps of (a) the size $S$ of the largest component, (b) the absolute magnetization $|m|$ in the final state, and (c) the active-link density $\rho^\text{surv}$ in survival runs as a function of $p$ and $q$ on random regular networks.
In the color map of $\rho^{\text{surv}}$, a value of $\rho^{\text{surv}}=0$ is specifically assigned to red to  distinguish from nonzero $\rho^{\text{surv}}$.
The white dashed line represents the percolation threshold $q_c=1/3$ of the fixed-link subgraph.
Panels (d)--(f) show the corresponding quantities as functions of $p$ for selected values of $q$. The solid lines are visual guides.
}
\end{figure*}

To validate our theoretical analysis, we implement numerical simulations of the CVM on random regular networks with fixed links.
We fix the degree $z=4$ and the system size 
$N=4\times10^3$ unless otherwise stated. 
Initially, each node is assigned one of the two states, $\sigma_i=+1$ or $-1$, 
with equal probability, and a fraction $q$ of links 
is randomly selected as fixed links.

To characterize the behavior of the model, we measure the size $S$ of the largest connected component and the absolute magnetization $|m|$, where $m=\frac{1}{N}\sum_{i=1}^{N}\sigma_i$. An active phase of the system is associated with a
finite value of the density $\rho$ of active links, but in a finite system an absorbing phase in which $\rho=0$ is reached due to finite-size fluctuations. Therefore, to account for an active phase in our simulations, we measure the density              $\rho^{\mathrm{surv}}$ of active links over surviving runs, that is, over realizations of the dynamics which,
at a given time, have not reached an absorbing state. Additional observables, including the activity inside and outside the fixed-link giant component and the size difference between the two largest components, are introduced when characterizing the localized active state.

Our numerical results confirm the existence of the three phases predicted in the phase diagram of our theoretical analysis. The analysis of Section III assumes the thermodynamic limit (infinite system) and we find here the finite-size system manifestation of these phases. Figures~\ref{fig_RR}(a)--(c) show the color maps of the
largest component size $S$, the absolute magnetization $|m|$, and
the density of active links in surviving runs, $\rho^{\rm surv}$  in the $(p,q)$ plane, which can also be interpreted as finite-size system phase diagrams. The active phase becomes, by finite-size fluctuations, a consensus phase with all nodes in the same state. This is characterized by $S=1$ and  $|m|=1$. The non-zero value of  $\rho^\text{surv}$ in this phase indicates the activity of the phase in the steady-state in the thermodynamic limit.

For small $q$, the system exhibits a conventional fragmentation transition from the consensus phase into a fragmented phase. The latter is the finite-size manifestation of the frozen phase. In this transition, $S$ and $|m|$ decrease from nearly unity to approximately $1/2$ and $0$, respectively, as $p$ increases and the system enters the fragmented phase.
Correspondingly, $\rho^{\text{surv}}$ decreases to zero.
As the fraction of fixed links increases, the fragmentation transition shifts toward a larger $p$, 
in agreement with the theoretical prediction.
When $q$ exceeds the percolation threshold $q_c$, the giant component is formed by
fixed links, leading to an asymmetric fragmented phase with $S>1/2$ and $|m|>0$ even for large $p$. This phase is the finite-size manifestation of the localized active phase found in the thermodynamic limit, and therefore $\rho^{\text{surv}}$ remains nonzero.
For sufficiently large $q$, the largest component remains close to the system size regardless of $p$.

\subsection{Delay of the fragmentation phase transition}

We now examine how the fraction of fixed links affects the
conventional fragmentation transition of the CVM. Figures~\ref{fig_RR}(d)--(f) show $S$, $|m|$ and $\rho^{\text{surv}}$ as
functions of the rewiring probability $p$ for several values of $q$.
The impact of fixed links systematically shifts this transition to 
larger values of $p$. As $q$ increases, a larger fraction of links are 
not available for rewiring, reducing the effective rewiring probability.
Consequently, stronger rewiring is required to overcome the
copying dynamics, leading to the delay of the fragmentation transition.
Therefore, the transition point moves to larger values of the plasticity $p$ when increasing $q$, as predicted 
in the moment-closure approximation. 
The same trend is clearly observed in $\rho^{\rm surv}$ shown in
Fig.~\ref{fig_RR}(f). For small $q$, $\rho^{\rm surv}$ decreases to
zero as $p$ increases, indicating the fragmentation phase transition.
With increasing $q$, the disappearance of $\rho^{\rm surv}$ is shifted 
to larger $p$.

\begin{figure}[t]
\centering
\includegraphics[width=\linewidth]{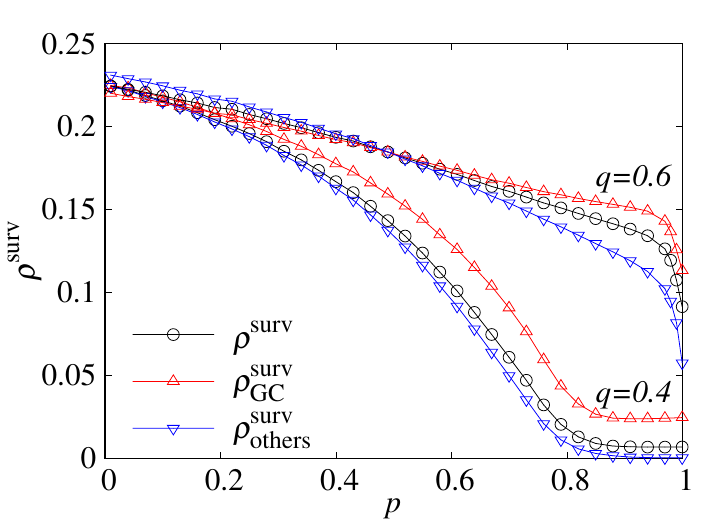}
\caption{\label{fig_localization}
The active-link density in survival runs in the entire network $\rho^{\rm surv}$, inside the fixed-link giant component $\rho^{\rm surv}_{\rm GC}$, and outside the fixed-link giant component $\rho^{\rm surv}_{\rm others}$ as a function of $p$ for $q=0.4$ and $0.6$ on the random regular network with $N=4000$ and $z=4$. The solid lines are visual guides.
}
\end{figure}

\begin{figure*}[t]
\centering
\includegraphics[width=\linewidth]{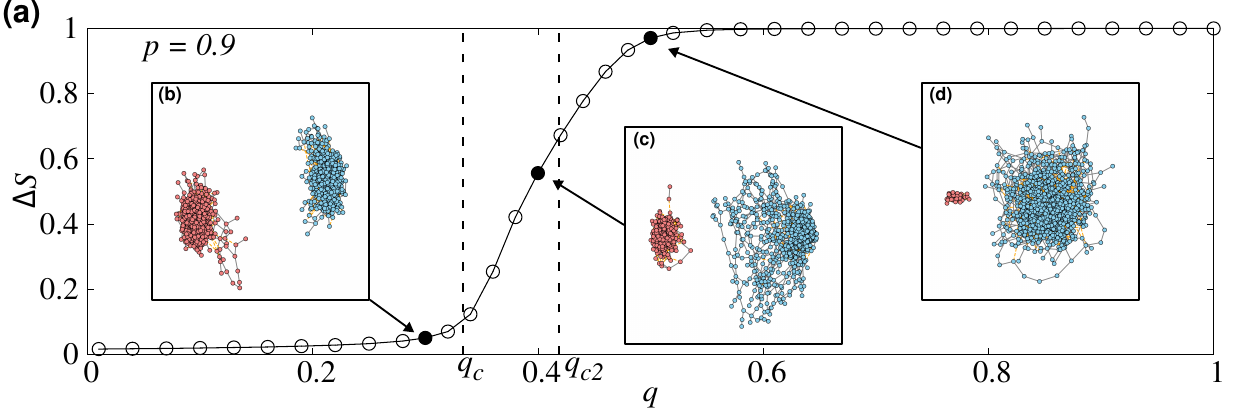}
\caption{\label{fig_asymmetry}
(a) The difference in the sizes of the largest (major state) and the second-largest (minor state) connected components $\Delta S$ at the steady state  as a function of $q$ on random regular networks with $N=4000$ and $z=4$. Here, the rewiring probability is set to $p=0.9$. The solid line connecting symbols is a visual guide, and two dashed lines represent $q_c=1/3$ and the location of the absorbing phase transition $q_{c2}$ predicted by the triplet approximation at $p=0.9$.
(b)--(d) Network configurations with $N=1000$ (for visual clarity) in the final state for $q=0.3, 0.4$ and $0.5$, respectively. The points corresponding to the configurations are marked by filled circles in (a).
}
\end{figure*}

\subsection{Percolation of fixed links and localized active state}

We observe that when increasing the plasticity $p$ for sufficiently large $q$ the fragmentation transition discussed above is not found. Rather, a finite surviving activity remains even at large $p$, indicating that the system no longer exhibits a frozen state. The disappearance of the frozen phase and the change in the dynamics at large $q$ are associated with the percolation of the fixed-link subgraph and the emergence of a localized active state. 
Indeed, as discussed in Sec.~IIIA, the subgraph composed solely
of fixed links undergoes a bond percolation transition at $q_c$, 
i.e., $q_c=1/3$ in our example.
Below this threshold, the fixed links form only finite connected components.
Therefore, when $p$ exceeds the absorbing transition threshold, rewiring 
separates nodes with different opinions into disconnected, internally 
ordered components and the system reaches the frozen fragmented state. 
In contrast, for $q>q_c$, the fixed
links form a giant connected component that remains intact throughout
the dynamics. This permanent backbone prevents complete structural
fragmentation and provides a connected region in which different
opinions can remain dynamically active.

To examine whether the surviving activity is indeed associated with
this fixed-link giant component, we separately measure the active-link
density inside and outside the giant component formed by fixed links.
Figure~\ref{fig_localization} shows the total surviving active-link
density $\rho^{\rm surv}$ together with the corresponding densities
inside the fixed-link giant component, $\rho^{\rm surv}_{\rm GC}$, and 
outside it, $\rho^{\rm surv}_{\rm others}$.
For $q=0.4$, slightly above the percolation threshold $q_c$,
$\rho^{\rm surv}_{\rm GC}$ remains finite at large $p$, whereas
$\rho^{\rm surv}_{\rm others}$ is nearly zero. 
This shows that the surviving activity is localized within 
the fixed-link giant component.
For the larger value $q=0.6$, both $\rho^{\rm surv}_{\rm GC}$ and
$\rho^{\rm surv}_{\rm others}$ remain non-zero and finite even at large $p$, indicating that the activity is sustained throughout the network.
These results show the distinction between globally active and localized active phase.
In addition, these results confirm the emergence of a localized active state
associated with the percolation of fixed links. Above $q_c$, the
percolating fixed backbone can sustain activity even under strong
rewiring, while activity outside the backbone is strongly suppressed.
The localized active phase is therefore distinct from both the
globally active phase and the frozen fragmented phase of the
conventional CVM.

\subsection{Asymmetric fragmentation in finite systems}

In a finite system, the localized active state  eventually reaches an absorbing state driven 
by finite-size fluctuations.
Its finite-size signature is an asymmetric fragmented state rather than 
the nearly symmetric fragmentation typically observed in the conventional CVM.
Because the fixed-link giant component provides a macroscopically connected backbone, 
one of the final components tends to remain larger than the others. 
Thus, the asymmetric fragmentation can be regarded as the manifestation in finite systems 
of the localized active phase.

To characterize the asymmetric fragmentation, we measure 
the difference between the sizes of the largest and the second largest 
connected components, $\Delta S=S_1-S_2$, at the final state with $p=0.9$ (Fig.~\ref{fig_asymmetry}).
For small $q$, the final fragmented state is nearly symmetric and
$\Delta S$ remains close to zero. Once the fixed links form a giant 
component at $q_c$, the fragmentation becomes increasingly
asymmetric. 
It results in a finite $\Delta S$, indicating asymmetric fragmentation for $q_c<q \lesssim q_{c2}$, where $q_{c2}$ is the critical fraction of fixed links obtained by solving $p_c(q_{c2})=0.9$ predicted by the triplet approximation.
In our example, $q_c=1/3$ and $q_{c2} \approx 0.42$.
For larger $q$, i.e., $q \gtrsim q_{c2}$, the system becomes the globally active state, leading to $\Delta S$ approaching unity. 
This behavior provides a direct measure of the asymmetric
fragmentation associated with the localized active state.

\section{Discussion}

We have studied the coevolving voter model in the presence of fixed links that 
are unbreakable. Fixed links modify the dynamics in two distinct ways: 
First, they reduce the effective rewiring process and shift the absorbing 
transition toward larger rewiring probabilities. 
Second, the subgraph formed by fixed links undergoes a percolation transition 
and generates a permanent giant component.
The combination of these dynamical and structural effects gives rise
to a new localized active state, in addition to the globally active and frozen
phases as in the conventional CVM.
In the localized active state, strong rewiring suppresses activity in
most of the network, while active links persist predominantly within
the giant component formed by fixed links. 
We also found that in any finite system, fluctuations eventually terminate
this active state, resulting in an asymmetric fragmented configuration
in which one component is substantially larger than the others.
Together with the moment-closure approximation and the percolation analysis, 
our results show that structural constraints can change not only 
the location of an absorbing transition but also the topological
organization of the surviving activity in coevolving networks.
Our findings suggest that persistent interactions in coevolving systems can play an important role in determining macroscopic states. This observation motivates further investigation into the role of persistent interactions. One natural direction for future work is to investigate the effects of fixed-link topology. While this study focuses on a random distribution of fixed links, non-random organization of fixed links, such as degree-dependent placement~\cite{moreira2009degree,shiraki2010target}, may significantly influence the formation and competition of macroscopic states.
Another interesting direction is to consider both fixed links and fixed nodes, called zealots~\cite{kitching2025breaking}, in adaptive dynamics.

\appendix

\section{Moment-closure approximation}
\label{app:moment_closure}

Here we provide the details of the moment-closure approximation used
to obtain the stationary active-link density and the absorbing
transition. We consider the symmetric state and employ a homogeneous
approximation in which the fixed-link disorder is treated in an
annealed manner. Accordingly, an active link is effectively subject
to rewiring with probability $p(1-q)$ and to copying with probability
$1-p(1-q)$.

We denote by $[AB]$ the number of active links per node. 
For a network with mean degree $z$, the density of active links is
\begin{align}
\rho=\frac{2[AB]}{z}.\label{app:rho_AB}
\end{align}
The evolution equation for $[AB]$ is
\begin{align}
\frac{d[AB]}{dt}&=\frac{2}{z}\Big\{[1-p(1-q)]\big[(z-2)[AB]-4[ABA]\big] \nonumber \\
&-p(1-q)[AB]\Big\}.
\label{app:dAB}
\end{align}
Here $[ABA]$ denotes a triplet in which the central node is connected
to two nodes in the opposite state. We have used the symmetry $[ABA]=[BAB]$.
The evolution of the triplet density involves quadruplet motifs.
Following the second-order moment expansion for the coevolving voter
model \cite{demirel2014moment}, its evolution can be written as
\begin{widetext}
\begin{align}
\frac{d[ABA]}{dt}=\frac{1}{z}\Big\{[1-p(1-q)]\big[-4[ABA]+2[ABBA]+[(A)BAA]-[ABAB]-3[(B)AAA]\big]
-4p(1-q)[ABA]\Big\}.
\label{app:dABA}
\end{align}
\end{widetext}
Here $[XYZW]$ denotes a chain quadruplet, whereas $[(X)YZW]$
denotes a star quadruplet with the state $X$ at the central node.
The appearance of quadruplet densities in Eq.~(\ref{app:dABA})
reflects the hierarchy of moment equations.

To close the hierarchy at the triplet level, we express the
quadruplet densities in terms of lower-order motifs. 
For the chain quadruplets, we use
\begin{align}
[ABAB]&\simeq\frac{4[ABA][BAB]}{[AB]}=\frac{4[ABA]^2}{[AB]},
\label{app:ABAB}
\\
[ABBA]&\simeq\frac{[ABB]^2}{4[BB]}.
\label{app:ABBA}
\end{align}
In the symmetric homogeneous state, the remaining pair and triplet
densities can be expressed as
\begin{align}
[AA]&=[BB]=\frac{z}{4}-\frac{[AB]}{2},
\label{app:AA}
\\
[AAB]&=[ABB]=(z-1)[AB]-2[ABA].
\label{app:AAB}
\end{align}
For a fixed coordination number $z$, the corresponding star
quadruplet closures are
\begin{align}
[(A)BAA]&\simeq\frac{z-2}{2(z-1)}\frac{[AAB]^2}{[AB]},
\label{app:star1}
\\
[(B)AAA]&\simeq\frac{2(z-2)}{3(z-1)}\frac{[ABA]^2}{[AB]}.
\label{app:star2}
\end{align}
The factors $(z-2)/(z-1)$ account for the finite number of remaining
neighbors once one edge incident to the central node has already
been specified.

Substituting Eqs.~(\ref{app:ABAB})--(\ref{app:star2}) into Eq.~(\ref{app:dABA}) 
closes the moment hierarchy in terms of$[AB]$ and $[ABA]$.
At the stationary state,
\begin{align}
\frac{d[AB]}{dt}=\frac{d[ABA]}{dt}=0.
\end{align}
From Eq.~(\ref{app:dAB}), the stationary triplet density can 
be expressed in terms of the density of $[AB]$ as
\begin{align}
[ABA]=\frac{(1-\tilde p)(z-2)-\tilde p}{4(1-\tilde p)}[AB],
\label{app:ABA_AB}
\end{align}
where $\tilde p=p(1-q)$.
Substituting Eq.~(\ref{app:ABA_AB}) together with the closure
relations into Eq.~(\ref{app:dABA}) and solving for $[AB]$ gives
\begin{align}
[AB]^*=\frac{z}{4}\frac{(z-1)^2\tilde p^{\,2}-2(z^2-2z+2)\tilde p+z(z-2)}
{(z-1)^2\tilde p^{\,2}-(2z^2-3z+2)\tilde p+z(z-1)}.
\label{app:AB_stationary}
\end{align}
Using Eq.~(\ref{app:rho_AB}), we obtain the steady state solution of 
the density of active links:
\begin{align}
\rho^*=\frac{(z-1)^2\tilde p^{\,2}-2(z^2-2z+2)\tilde p+z(z-2)}
{2\left[(z-1)^2\tilde p^{\,2}-(2z^2-3z+2)\tilde p+z(z-1)\right]}.
\label{app:rho_stationary}
\end{align}

The absorbing transition occurs when the non-trivial stationary
solution disappears, i.e., when $\rho^*=0$. 
Thus, the numerator of Eq.~(\ref{app:rho_stationary}) satisfies
\begin{align}
(z-1)^2\tilde p_c^{\,2}-2(z^2-2z+2)\tilde p_c+z(z-2)=0.
\label{app:critical_poly}
\end{align}
The physically relevant solution is then
\begin{align}
\tilde p_c&=\frac{\sqrt{3z^2-6z+4}-2}{\sqrt{3z^2-6z+4}+1}.
\label{app:ptilde_c}
\end{align}
Finally, using $\tilde p_c=p_c(1-q)$, we obtain
\begin{align}
p_c(q)=\left(\frac{1}{1-q}\right)\frac{\sqrt{3z^2-6z+4}-2}{\sqrt{3z^2-6z+4}+1},
\end{align}
which is Eq.~(\ref{eq:pc}) in the main text.

\begin{acknowledgments}
This research was supported in part by the National
Research Foundation of Korea (NRF) grant funded by
the Korea government (MSIT) (No. RS-2025-25433094)
and by Global - Learning \& Academic research institution 
for Master’s $\cdot$ PhD students, and Postdocs (LAMP)
Program of the National Research Foundation of Korea
(NRF) grant funded by the Ministry of Education (No.
RS-2024-00445180). MSM acknowledges partial support from Grant No. PID2024-157493NB-C21 funded by MI-CIU/AEI/10.13039/501100011033 and “ERDF/EU”, and from Grant No. CEX2021-001164-M funded by the Maria de Maeztu Program for Units of Excellence in R$\&$D
\end{acknowledgments}

\bibliography{cvm}

\end{document}